\documentclass{article}
\usepackage{amsfonts, epsfig}
\usepackage{amssymb}

\newcommand{\dd} {\mbox{d\raisebox{0.75ex}{\hspace*{-0.32em}-}\hspace*{-0.02em}}}

\title{Eta Production in the pp Scattering}
\author{Sa\v sa Ceci and Alfred \v Svarc \\ Ru\dd er Bo\v skovi\' c Institute, Bijeni\v
cka Cesta 54, Zagreb, Croatia}

\def\Journal#1#2#3#4{{#1} {\bf #2}, #3 (#4)}

\def\NPA{{\em Nucl. Phys.} A}
\def\PLB{{\em Phys. Lett.} B}

\def\PRC{{\em Phys. Rev.} C}

\def\PS{\em Physica Scripta}

\def\be{\begin{equation}}
\def\ee{\end{equation}}
\def\bea{\begin{eqnarray}}
\def\eea{\end{eqnarray}}

\begin{document}
\maketitle

\section{Introduction}

Zagreb-UCLA-ANL meson nucleon partial wave analysis\cite{Bat95}
opened the possibility to explain the number of intermediate
energy data. Important result of the PWA are the eta nucleon
elastic T matrices. The three-body processes can serve us to lower
the error of the two-body T matrices. The idea is to take two-body
T matrices from existing PWA's and apply them to the three-body
processes.

The most convenient process for that purpose is the $pp\rightarrow
pp\eta$. It is the isospin 1/2 process, what enables us to test
the half of the amplitudes at a time. Dominant contribution to the
production process comes through meson-exchange\cite{Machleidt},
and decay of nucleon resonances\cite{Bat97,Vet91,RhoM}.

\section{Model}
Our model has been described earlier\cite{Bat97}. Because of the
large uncertainties in the treatment of final state interaction of
the considered theoretical models\cite{Vet91,RhoM}, we have
decided to ignore its influence at the moment. The FSI is
significant near threshold, so energies from 100 to 500 MeV would
be the most convenient place to test this model. Unfortunately, in
this range, experimental data are very hard to get. Preliminary
results are given in the Figure 1.

\section{Discussion}

Light and dark lines show constructive and destructive total cross
section, respectably. Enormous discrepancy below few MeV are
expected because we have completely disregarded final state
interaction. Our preliminary results again do not show rho meson
dominance - which would be expected from most of the other
models\cite{RhoM}. To get better picture about what is going on,
we shall calculate further differential cross sections, and total
cross sections for pn to $\eta$pn. Completed results will be
reported elsewhere.

\begin{figure}[h]
\begin{center}
\vspace{33pt} \epsfig{figure=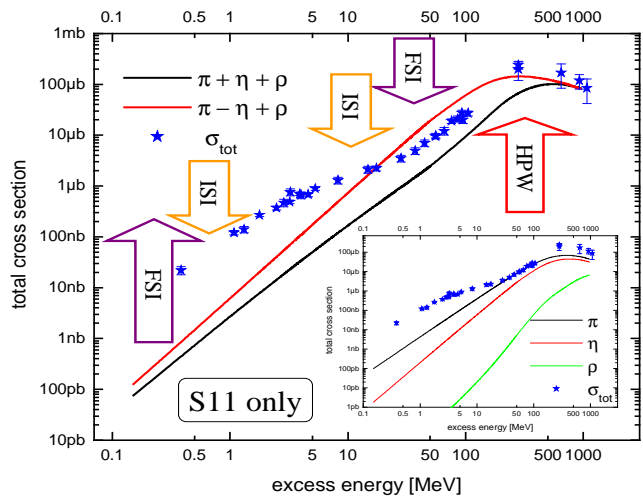,height=6.8cm,width=8.8cm}
\vspace{-3pt}
\caption[]{Experimental data are from
\protect\cite{ExpD}. Figure shows, qualitatively, not bad behavior
of constructive sum of $\pi$ and $\eta$ contributions. Moreover,
added corrections of initial and final interaction will push
theoretical curve in the right direction. In the small graph is
given contributions for each considered meson exchange.}
\end{center}
\end{figure}


\begin{thebibliography}{9}

       \bibitem{Bat95} Batini\' c, M. {\em et al}: \Journal{\PRC}{81}{2310}{1995}. \\
        Batini\' c, M. {\em et al}: \Journal{\PS}{58}{15}{1998}.

       \bibitem{Machleidt} Machleidt R.: {\it Adv. Nucl. Phys.} {\bf 19}, 189 (1989).


       \bibitem{Bat97} Batini\' c, M., \v Svarc, A. and Lee,
       T.-S. H.:
        {\it Physica Scripta} {\bf 56}, 321 (1997).

      \bibitem{Vet91}
       Vetter, T., Engel, A., Bir\' o, T. and Mosel, U.:
       \Journal{\PLB}{263}{153}{1991}.

      \bibitem{RhoM}
       Laget, J. M. and Wellers, F.: \Journal{\PLB}{657}{254}{1991}. Gedalin, E., Moalem, A. and Razdolskaja,
       L.:
       \Journal{\NPA}{634}{368}{1998}. Santra, A. B. and Jain, B. K.:
       \Journal{\NPA}{634}{309}{1998}. Hanhart, C. and Nakayama, K.:
       \Journal{\PLB}{454}{176}{1999}.

       \bibitem{ExpD}
       Cal\' en, H. {\em et al}: \Journal{\PLB}{458}{190}{1999}.\\
       Moskal, P. {\em et al}:
       \Journal{\PLB}{482}{356}{2000}.


\end{thebibliography}
\end{document}